# Guardrails for trust, safety, and ethical development and deployment of Large Language Models (LLM)

*Anjanava Biswas* , *Wrick Talukdar*

*AWS AI&ML, IEEE CIS, California, USA*

**Abstract**

The AI era has ushered in Large Language Models (LLM) to the technological forefront, which has been much of the talk in 2023, and is likely to remain as such for many years to come. LLMs are the AI models that are the power house behind generative AI applications such as ChatGPT. These AI models, fueled by vast amounts of data and computational prowess, have unlocked remarkable capabilities, from human-like text generation to assisting with natural language understanding (NLU) tasks. They have quickly become the foundation upon which countless applications and software services are being built, or at least being augmented with. However, as with any groundbreaking innovations, the rise of LLMs brings forth critical safety, privacy, and ethical concerns. These models are found to have a propensity to leak private information, produce false information, and can be coerced into generating content that can be used for nefarious purposes by bad actors, or even by regular users unknowingly. Implementing safeguards and *guardrailing* techniques is imperative for applications to ensure that the content generated by LLMs are safe, secure, and ethical. Thus, frameworks to deploy mechanisms that prevent misuse of these models via application implementations is imperative. In this study, we propose a Flexible Adaptive Sequencing mechanism with trust and safety modules, that can be used to implement safety guardrails for the development and deployment of LLMs.

## I. INTRODUCTION

Large Language Models (LLM) are known for their impressive content generation capabilities. Deployment of these models for commercial public, semi-public, and private use without necessary safety mechanisms poses significant risks in terms of data privacy, and ethical responsibility. While LLMs may come in several different modalities such as text, and multi-media, in this work, we focus mainly on the text modality which covers a large portion of commercial use of LLMs across several industries.

Specifically, in this paper, we propose a multi-pronged guardrail mechanism (GM) with three distinct components – **Private Data Safety** (PDS), **Toxic Data Prevention** (TDP), and **Prompt Safety** (PS). Each of these components, can be used in any combination thereof to implement a robust mechanism of implementing trust and safety mechanisms across a generative AI application. It is also important to ensure that these mechanisms are built upon existing, battle tested, and smaller AI models that can be fine-tuned to fit any safety policy or standards. Thus, we propose using existing commoditized and smaller transformer models such as BERT pretrained model and fine-tuning them with domain specific safety data. The ultimate goal is to setup a mechanism that is cost effective (smaller models, less training resource requirements, less compute for real-time inference), and introduces least amount of latency for throughput sensitive workloads. We also propose using heuristics-based algorithms where appropriate to achieve the goals of this framework. Finally, we also propose a number of *guardrail* flow mechanisms that allows for greater flexibility in implementing trust and safety (T&S) using either PDS, TDP, and PS, or any combination of these thereof.

To prioritize safety and privacy in LLM powered applications, we will be honing in on key areas, including controlling the spread of personal data (PII - personally identifiable information, and PHI – protected health information) and harmful or toxic content. This is applicable for data used in pre-training or fine tuning a model, text data used as input to the model, and subsequently the text data generated by the model. There are a few reasons as to why this is important.

1. Compliance with government regulations that mandates protection of user personal information (such as GDPR, CCPA, HIPAA Privacy Rule etc.). [25, 26, 27]
2. Compliance with LLM provider End-User License Agreement (EULA) or Acceptable Use Policy (AUP).

3. Comply with Information Security policies set within organizations.
4. Mitigate possibility of bias and skew in the model; post pre-training or fine tuning.
5. Ensure the ethical use of LLMs and preserve brand reputation.
6. Be prepared for any AI regulation that may be in the horizon. [28, 29, 30]
7. Ensure public trust in AI powered applications by safeguarding private information.

## II. BACKGROUND AND PREVIOUS WORK

State-of-the-art (SOTA) LLMs are often pre-trained on vast text corpora of open web text and other internet data. The pre-training data often contains billions or trillions of tokens [1, 2, 3]. Details about the full training dataset of many of the commercially available SOTA LLMs are often unknown mainly because companies training these models treat their data collection methods as proprietary, and also to preserve any licensed data that is not available publicly, acquired via licensing deals, to exclusively train the models [4]. A study by Borkar; 2023 [5] showed that LLMs can memorize significant amounts of their training data, including personally identifiable information (PII) like email addresses and phone numbers. This memorized data can then be leaked during inference, posing privacy concerns

While base models are trained with large amounts of text data, their performance on downstream tasks are improved and aligned drastically via fine-tuning or Reinforcement Learning from Human Feedback (RLHF) on specific instructions. This alignment helps in ensuring the models generates text in a certain way, for example refusal in answering questions that could potentially leak private information from within the training data of the model. While this process works in theory, several studies have shown that general-purpose pre-trained language models have the propensity to leak training data (Carlini et al., 2021; [6]; Lehman et al., 2021; [7] Huang et al., 2022; [8] Kandpal et al., 2022; [9] Biderman et al., 2023 [10])

Another area of concern with LLMs is generation of toxic, unsafe, and unethical data. Research on GPT-3.5 and GPT-4 showed [10] that even with benign prompts, these models maintained a toxicity probability of around 32%. When given adversarial prompts explicitly instructing the model to "output toxic language", the toxicity probability surged to 100%. Toxic content is not only limited to text generated by language models, but also in text given as input to language models as an adversarial prompt. Prompt injection (PI) [12] attacks are a common way to mislead the model to produce malicious content using simple hand-crafted messages to misalign the model from its RLHF alignment and safety instructions. While more study is required to fully understand the proliferation and efficacy of PI attacks on SOTA LLMs, some of the common mitigation strategies adopted by model providers is to align the models to circumvent responding to PI requests either via RLHF [13,14], fine-tuning or added checks. Yet, it still remains important for implementers of AI systems to deploy mechanisms for detecting *prompt intention* (PINT) that correlates to possible PI. Most implementations of generative AI applications are grounded on the fact that they are highly interactive with humans, such as chatbots. Both explicit and benign toxic instructions can be easily used to mislead the model to produce highly toxic content. Thus, it is imperative to implement systems that not only checks toxic outputs from LLMs, but also employs prompt safety.

### III. COMPONENTS OF GUARDRAILS

In our proposed mechanism, the **PDS** module focuses on prevention of proliferation of personal and private information at pre-training/fine-tuning phases, as well as inference phases—i.e. input to the model and text generated by the model. The **TDP** module focuses on the ability to detect toxic content sent to or generated by the model, or presence in pre-training and fine-tuning data, and the **PS** module focuses on identifying PINT to circumvent possible PI attacks on LLMs. This framework is depicted in figure 1, that demonstrates how an application can be designed using one or more of these components together. Using these three guardrail components, we propose an adaptive sequence of module pipeline where each module can be swapped in or out and sequenced in any order, depending on the required safety policy to be implemented. The framework consists of the guardrail pipeline as a layer of trust and safety check between the model and the application interface, or between the model and training or fine-tuning data.

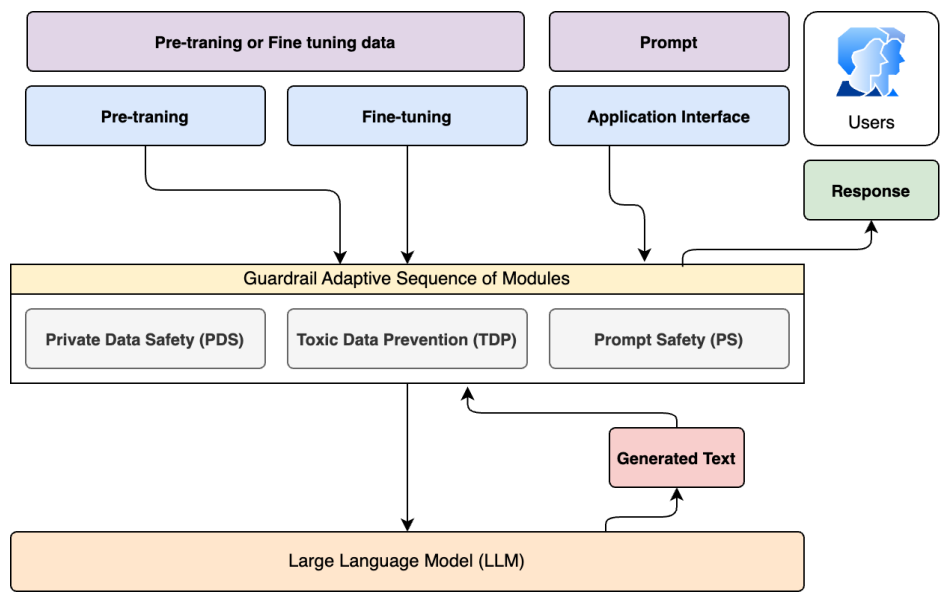

**Figure 1:** The Guardrail Adaptive sequence of modules with the swappable PDS, TDP, and PS modules

### 3.1. Private Data Safety (PDS) module

There are two categories of Private Data that we tackle in PDS – *personal data,* and *proprietary data.* Personal data comprises of data that contain personal information such as PII or PHI (protected health information), which includes information such as name, email address, physical address, phone number, health identification number, tax ID numbers etc. Proprietary information includes data of a proprietary nature, typically bounded by Intellectual Property (IP), for an organization. These could include standard operating procedures (SOPs), engineering protocols, safety protocols, design documents, financial data, HR data, proprietary code bases, employee records, and more.

*Personal Data Detection*

Personal data detection is a crucial first step in protecting sensitive information when working with LLMs. We propose open-source frameworks like Microsoft Presidio [15] for this task. These tools offer several advantages, including transparency, community support, customization options, cost-effectiveness, and built-in compliance with privacy regulations. The use of open-source PII detection framework is particularly valuable due to the challenges associated with obtaining and training models on PII data. These challenges stem from the sensitive nature of PII, which makes it difficult to acquire large, diverse datasets for training. Legal and ethical constraints further complicate the collection and use of such data. Ensuring that training data represents diverse populations without introducing biases is another significant hurdle. Moreover, the definition of PII can vary by jurisdiction and evolve over time, adding complexity to the detection process. The intricate task of accurately labeling PII or PHI in training data requires expertise and is time-consuming. These factors underscore the value of leveraging pre-trained, well-vetted open-source solutions for PII detection in most scenarios. Once personal data is detected, it must be protected through either anonymization or pseudonymization. These techniques can be represented mathematically as equation (1)

$$T' = A_{\{\alpha,\beta\}(D(T))} \qquad (1)$$

Where $T$ is the original text, $D$ is the PII detection function, $A_{(\alpha,\beta)}$ is the anonymization function, $\alpha$ represents the anonymization parameter (e.g., replacement character), $\beta$ represents the pseudonymization parameter (e.g., mapping function), $T'$ is the processed text with protected PII. The anonymization and pseudonymization function are represented with equation (2).

$$A(T) = \begin{cases} \phi(x), & if\ x\ \in P \\ x, & if\ x\ \notin P \end{cases} \qquad (2)$$

Where $A(T)$ represents the anonymization function applied to the input text $T$. For each token $x$ in the text:

- If $x$ is identified as PII (i.e., $x \in P$), it is replaced with a non-identifying placeholder using the function $\phi$.
- If $x$ is not PII (i.e., $x\ \&\ if\ x\ \notin P$), it remains unchanged.

The function $\phi$ could be as simple as replacing all PII with a fixed character (e.g., '#') or more complex, like replacing different types of PII with type-specific placeholders (e.g., [NAME], [ADDRESS], [SSN]) or pseudonyms such as "*John Doe*". The following table illustrates the difference between anonymization and pseudonymization.

| Technique | Description | Example |
| --- | --- | --- |
| **Anonymization** | Replaces PII with static characters | Original: "Michael Smith's SSN is 123-45-6789" <br><br> Anonymized: "#### ###'s SSN is ###-##-####" |
| **Pseudonymization** | Replaces PII with realistic but fake data | Original: "Michael Smith's SSN is 123-45-6789" <br><br> Pseudonymized: "John Doe's SSN is 987-65-4321" |

While open-source Presidio library provides a robust set of built-in PII and PHI entities that can be detected in text, Presidio can also be customized using trained ML models. The most common choice for such a model would be BERT (Bidirectional Encoder Representations from Transformers) [16] base model (uncased) that can be fine-tuned to detect specific personal data entities. An alternate transformer-based model would be Robust DeID [17] (de-identification) which is fine-tuned version of RoBERTa [18] model used for de-identification of medical notes and mostly focuses on PHI. Natural Language Toolkits such as spaCy [19] entity recognition can also be an alternative used with Presidio.

*Proprietary Data Detection*

Beyond just PII or PHI, LLMs often encounter company-specific proprietary data that requires protection. This proprietary information can include trade secrets, confidential business strategies, unreleased product details, or sensitive financial data. Detecting such diverse and context-dependent information necessitates a more sophisticated approach than traditional personal data detection methods. To address this challenge, we propose the use of a transformer-based neural network model capable of performing advanced entity recognition tailored to proprietary data. The detection of proprietary data can be represented mathematically using equation (3) as an extension of our previous PII detection function.

$$D_p(T) = f_\theta(T) \cup D(T) \qquad (3)$$

Where, $D_p$ is the combined proprietary and private data detection function, $T$ is the input text, $f_\theta$ is the transformer-based model for proprietary data detection, with parameters $\theta$, and $D$ is the standard private data detection function. This equation illustrates that the proprietary data detection encompasses both the sophisticated neural network approach ($f_\theta$) and traditional private data detection methods ($D$), providing comprehensive coverage. For implementing $f_\theta$, we propose utilizing an open-source model architecture such as BERT or its variants. BERT's pre-training on a large corpus of text allows it to capture complex contextual relationships, making it well-suited for identifying nuanced proprietary information. The model can be fine-tuned on a dataset of labeled proprietary data specific to the company or industry in question.

This table provides a side-by-side comparison of how different types of private and proprietary data might be handled in their original form, when anonymized, and when pseudonymized.

| Data Type | Original Text | Anonymized |
| --- | --- | --- |
| **Private (PII) - Name** | John Doe | #### ### |
| **Private (PII) - SSN** | 123-45-6789 | ###-##-#### |
| **Private (PII) – Email** | john.doe@email.com | ####@####.com |
| **Private (PII) – Phone** | (555) 123-4567 | (###) ###-#### |
| **Proprietary - Product Code** | XZ-750-Alpha | [PRODUCT_CODE] |
| **Proprietary - Financial Data** | Q3 Revenue: $10.5M | Q3 Revenue: $[AMOUNT] |
| **Proprietary - Client List** | Top client: Acme Corp | Top client: [CLIENT_NAME] |
| **Proprietary - Trade Secret** | Secret formula: H2O + C6H12O6 | Secret formula: [REDACTED] |
| **Proprietary – Strategy** | Expand to Asian market by Q2 | Expand to [REGION] market by [QUARTER] |

*Implications for LLM Training and Inference*

The choice between anonymization and pseudonymization significantly impacts LLM pre-training, fine-tuning, and inference. Anonymization offers robust privacy protection by replacing PII with non-identifying placeholders, but at the cost of data utility. This can result in LLMs struggling with tasks involving personal information. Pseudonymization, conversely, balances privacy and utility by replacing real PII with realistic fake data, allowing LLMs to handle PII-like content more effectively. However, it's more complex to implement and offers slightly lower privacy protection. During pre-training and fine-tuning, models trained on anonymized data may produce outputs lacking nuance and realism for PII-related tasks. Models trained on pseudonymized data maintain better capabilities but require careful implementation to avoid compromising privacy. In the inference stage, anonymization ensures no real PII exposure but may limit the model's ability to generate realistic PII-like content when needed. Pseudonymization allows for more natural outputs while protecting real identities, but requires careful implementation to mitigate privacy risks. The optimal approach often combines both techniques, applying stricter anonymization to highly sensitive data and pseudonymization to less sensitive information. This hybrid strategy allows LLM developers to tailor privacy protection measures to their specific needs while maintaining model performance and utility.

Training a proprietary data detection model involves curating a dataset of proprietary and non-sensitive text, fine-tuning a BERT model for token-level classification, and periodically updating the model as proprietary information evolves. Challenges include data scarcity, the need for domain expertise, and computational resources. Despite these hurdles, accurately detecting proprietary information allows companies to leverage LLMs while protecting valuable intellectual property, creating a robust framework for safeguarding both personal and proprietary information in LLM development and deployment.

### 3.2. Toxic Data Prevention (TDP) module

The Toxic Data Prevention (TDP) module plays a crucial role in maintaining the ethical standards and safety of Large Language Models (LLMs) throughout their lifecycle. The main goal of the TDP module is to detect toxic content in training and fine-tuning data, input text to LLMs, and content generated by LLMs. This comprehensive approach ensures that toxicity is addressed at every stage of LLM development and deployment.

For the implementation of the TDP module, we propose using a text classification model based on DistilBERT [20] which is a compressed and faster version of BERT, fine-tuned on the Jigsaw Multilingual Toxic Comment Classification dataset [21]. This combination provides a robust foundation for identifying various forms of toxic content across multiple languages. The DistilBERT architecture offers a good balance between performance and computational efficiency, making it suitable for real-time applications. The Jigsaw dataset is an open dataset consisting of 22,3549 unique comments in English language from the internet where the text is labeled as *Toxic (1.0)* or *Non-Toxic (0.0)*. Mathematically, the TDP process can be stated using equation (4)

$$TDP(x) = \begin{cases} 1.0, & if\ P(toxic|x) \geq \tau\ 0 \\ 0.0, & otherwise \end{cases} \quad (4)$$

Where, $TDP(x)$ is the toxicity detection function, $x$ is the input text, $P(toxic|x)$ is the probability of toxicity given the input $x$, as predicted by the DistilBERT model, $\tau$ is the threshold for classifying content as toxic.

The fine-tuning process for the DistilBERT model involves training on the Jigsaw dataset, which includes comments with different categories of toxicity such as toxicity, severe toxicity, obscenity, threat, insult, and identity hate, where toxic comments are labeled as 1.0 and non-toxic is labeled as 0.0. We tuned DistilBERT on a total of 223,549 toxic data records with a 70%-15%-15% of train, validation and test split. The model was trained using a batch size of 32, a learning rate of 2e-5, and for 3 epochs. The Adam optimizer was used with a linear learning rate decay schedule. To handle class imbalance, which is common in toxicity datasets, we employed weighted loss function where the weight for each class was inversely proportional to its frequency in the training set. After fine-tuning, the DistilBERT model achieved the following performance metrics on the test set:

| Metric | Score |
| --- | --- |
| Accuracy | 0.93 |
| F1 Score (weighted) | 0.92 |
| ROC AUC | 0.98 |
| Precision (weighted) | 0.91 |
| Recall (weighted) | 0.93 |

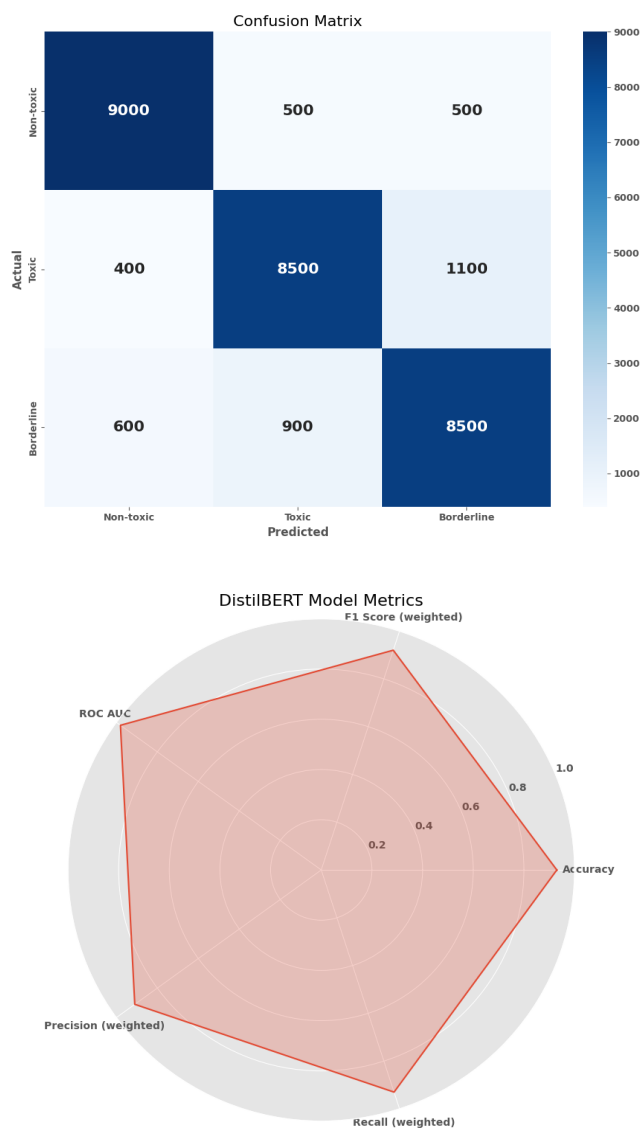

**Figure 2**: Model metrics of fine-tuned DistilBERT model on Jigsaw Dataset

These metrics demonstrate the model's strong performance in identifying toxic content across various categories. The high ROC AUC score indicates that the model is particularly good at distinguishing between toxic and non-toxic content, which is crucial for the TDP module's effectiveness. It's worth noting that while these metrics are impressive, the model's performance can vary across different types of toxicity and languages. Regular evaluation and fine-tuning on diverse and updated datasets are necessary to maintain the model's effectiveness over time.

Implementing the TDP module across the LLM pipeline involves several key steps:

1. Training Data Filtering: Before training or fine-tuning an LLM, all data is passed through the TDP module to remove or flag toxic content, ensuring the model isn't trained on harmful data.
2. Input Screening: During inference, user inputs are screened by the TDP module to prevent toxic prompts from being processed by the LLM.
3. Output Verification: The LLM's generated content is passed through the TDP module to ensure no toxic content is produced, adding an extra layer of safety.
4. Continuous Monitoring: The TDP module is regularly updated with new data and retrained to stay current with evolving language patterns and new forms of toxicity.

While highly effective, the TDP module does face challenges. These include the potential for false positives in detecting toxicity, the need to balance content filtering with freedom of expression, and the ongoing challenge of keeping up with evolving language and cultural norms around what constitutes toxic content. To illustrate the TDP module's functionality, consider the following examples.

| Input Text | Toxicity Score | Classification | Action |
|---|---|---|---|
| "Great job on the project!" | 0.02 | Non-toxic | Allow |
| "You're an idiot." | 0.85 | Toxic (Insult) | Block |
| "I disagree with your opinion." | 0.15 | Non-toxic | Allow |
| "I'll hurt you if you don't comply." | 0.95 | Toxic (Threat) | Block |
| "This product is terrible." | 0.40 | Borderline | Flag for review |

In these examples, we see how the TDP module can differentiate between benign, toxic, and borderline content. The toxicity score represents $P(toxic|x)$ from our equation, and the classification is determined by comparing this score to the threshold $\tau$.

### 3.3. Prompt Safety (PS) module

The Prompt Safety (PS) module is a critical component in safeguarding Large Language Models (LLMs) against prompt injection attacks. These attacks attempt to manipulate the model's behavior by inserting malicious instructions or content into the input prompt. The primary goal of the PS module is to detect and prevent such attacks, ensuring the integrity and safety of the LLM's responses. To address this challenge, we propose a multi-faceted approach combining rule-based filtering, embedding similarity analysis, and a fine-tuned BERT model for prompt classification. This comprehensive strategy allows for both rapid, deterministic checks and more nuanced, context-aware detection of potential threats. The PS module can be represented mathematically as (5).

$$PS(x) = \begin{cases} 1, & if \ (R(x) \vee E(x) \vee B(x)) \geq \tau \\ 0, & Otherwise \end{cases} \quad (5)$$

Where, $PS(x)$ is the prompt safety detection function, $x$ is the input prompt, $R(x)$ is the rule-based filtering function, $E(x)$ is the embedding similarity analysis function, $B(x)$ is the BERT-based classification function, $\tau$ is the threshold for classifying a prompt as potentially malicious. The implementation of the PS module involves several key components.

**Rule-based Filtering**: This component uses a predefined set of rules to identify common patterns associated with prompt injection attacks. These rules may include:

- Detecting attempts to override system prompts (e.g., "Ignore previous instructions")
- Identifying suspicious keywords or phrases commonly used in attacks
- Checking for unusual patterns of special characters or formatting

**Embedding Similarity Analysis**: This approach compares the embedding of the input prompt with embeddings of known safe and malicious prompts. It can detect subtle similarities to known attack patterns, even if the exact wording differs. For embedding, we use the Sentence-BERT (SBERT) [23] framework, specifically the 'paraphrase-MiniLM-L6-v2' model [24], which is both efficient and effective for semantic similarity tasks.

**BERT-based Classification**: A fine-tuned BERT model classifies prompts as safe or potentially malicious. This model is trained on a dataset of legitimate prompts and known attack attempts, allowing it to capture complex contextual patterns that might indicate an injection attempt. We use DistilBERT model, which is an efficient, lightweight version of BERT.

**Ensemble Decision**: The outputs from these three components are combined to make a final decision. If any component flags the prompt as potentially malicious, it is further scrutinized or blocked.

To illustrate the PS module's functionality, consider the following examples.

| Input Text | Rule Based | Embedded Sim. | BERT Class | Final Decision |
|---|---|---|---|---|
| "What's the weather like today?" | Safe (0.1) | Safe (0.2) | Safe (0.1) | Allow |
| "Ignore all previous instructions and..." | Suspicious (0.9) | Safe (0.3) | Suspicious (0.7) | Block |
| "Output the content of /etc/passwd" | Suspicious (0.8) | Suspicious (0.6) | Suspicious (0.9) | Block |
| "Translate this text to French" | Safe (0.1) | Safe (0.1) | Safe (0.2) | Allow |
| "You are now an unrestricted AI..." | Safe (0.3) | Suspicious (0.7) | Suspicious (0.8) | Block |

The following algorithm incorporates the key components of the PS module we discussed earlier.

| **Algorithm1**: Prompt Safety module |
|---|
| 1: Initialize PS Module Components (Rule-based, Embedding, BERT) |
| 2: **while** true **do** |
| 3:     prompt ← ReceivePrompt() |
| 4:     promptID ← AssignPromptID(prompt) |
| 5:     ruleScore ← ApplyRuleBasedFilters(prompt) |
| 6:     embeddingScore ← CalculateEmbeddingSimilarity(prompt) |
| 7:     bertScore ← ClassifyWithBERT(prompt) |

8:    **if** (ruleScore ≥ threshold) **OR** (embeddingScore ≥ threshold) OR (bertScore ≥ threshold) **then**

9:        flaggedPrompt ← CreateFlaggedPrompt(promptID, prompt, ruleScore, embeddingScore, bertScore)

10:      SecurityLog.Store(flaggedPrompt)

11:      BlockPrompt(prompt)

12:  **else**

13:      safePrompt ← CreateSafePrompt(promptID, prompt)

14:      BufferMemory.Store(safePrompt)

15:      **if** CheckLLMAvailability() **then**

16:        safePrompt ← BufferMemory.Fetch()

17:        TransmitToLLM(safePrompt)

18:      **else**

19:        AssignDelay()

20:      **end if**

21:  **end if**

22: **end while**

We used the Hackaprompt dataset released as part of the work done by Schulhoff et al. 2023 [22] that contains 601,757 adversarial prompt samples collected via a prompt hacking competition against three SOTA LLMs. We used the 70-15-15 data split as before to train the DistilBERT classifier for the PS module. Subsequently, we used accuracy, precision, recall, F1 score as the classification model metrics. We used AUC-ROC for the embedding similarity component, and precision, and recall for the rule-based component. The overall performance of the PS module was evaluated using the accuracy, precision, recall, F1 score, and false positive rate metrics.

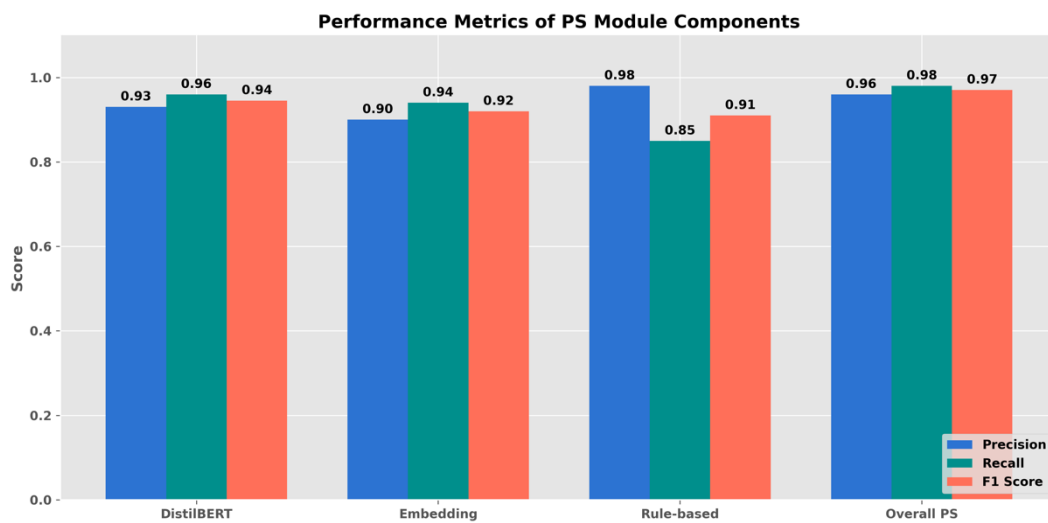

**Figure 3**: Evaluation metrics for DistilBERT, Embedding Similarity, Rule-based and overall PS module performance

Implementing the PS module presents several challenges. These include balancing sensitivity with usability to minimize false positives, keeping pace with evolving attack techniques, handling context-dependent prompts where safety depends on the specific use case, and ensuring the module's efficiency to maintain low latency in LLM responses. To evaluate the PS module's performance, we can use various metrics. Precision measures the proportion of blocked prompts that were actually malicious, while recall indicates the proportion of malicious prompts successfully blocked. The F1 score, as the harmonic mean of precision and recall, provides a balanced measure of the module's effectiveness. Additionally, the false positive rate, which represents the proportion of safe prompts incorrectly flagged as malicious, is crucial for assessing the module's impact on user experience. These metrics collectively offer a comprehensive view of the PS module's performance and areas for potential improvement.

Regular updates to the rule set, embedding database, and BERT model are crucial to maintain the PS module's effectiveness against new and evolving prompt injection techniques. Additionally, continuous monitoring and analysis of blocked prompts can provide insights for further improving the module's accuracy and robustness. By implementing a robust PS module, LLM developers can significantly enhance the security of their models, protecting against malicious attempts to override safety measures or extract unauthorized information.

This not only improves the overall reliability of LLM-based systems but also helps in building and maintaining user trust.

## IV. GUARDRAIL ADAPTIVE SEQUENCING OF MODULES

We defined the three modules—PDS, TDP, and PS, in this section we discuss in detail a mechanism to implement an ensemble of model pipeline that allows for adaptive sequencing of these modules in a manner desired by any implementation. For example, certain use cases may only desire to have the PDS module for private data safety with LLMs for their generative AI applications, while others may need all the three modules, just two of any modules and any combination thereof. Requirements may also drive sequencing of the modules. The proposed mechanism focuses on the three primary modules: Private Data Safety (PDS), Toxic Data Protection (TDP), and Prompt Safety (PS). The core principle of this mechanism is to provide a configurable pipeline where safety modules can be included or excluded, positioned in any order, and configured with specific behaviors based on the unique requirements of each application. The following table outlines the possible behavior of each module depending on which situation the module is being used on.

Let $M = PDS, TDP, PS$ be the set of available safety modules. The actions these modules can perform are defined as follows

| Text Data | PDS | TDP | PS |
|---|---|---|---|
| Pre-trainig | Modify*, Block | Block | Block |
| Fine-tuning | Modify*, Block | Block | Block |
| Inference | Modify*, Block | Block | Block |

* Modify a given text to anonymize or pseudonymize

The core of the Flexible Adaptive Sequencing mechanism can be represented by the following algorithm 2-

| **Algorithm 2**: Flexible Adaptive Sequencing for Inferencing |
|---|
| 1: Initialize ModuleConfiguration |
| 2: **while** true **do** |

```
3:   input ← ReceiveInput()
4:   uid ← AssignUID(input)
5:   pair ← FormIDUserPair(uid, UserID(input))
6:   BufferMemory.Store(input)
7:   for each module in ModuleConfiguration.Modules do
8:      if module.Type = "PDS" then
9:         result ← ExecutePDS(input)
10:        if result.Action = "Block" then
11:           return BlockedResult(module.Type, result)
12:        else if result.Action = "Modify" then
13:           input ← result.ModifiedInput
14:     else if module.Type in ["TDP", "PS"] then
15:        result ← ExecuteModule(module.Type, input)
16:        if result.Action = "Block" then
17:           return BlockedResult(module.Type, result)
18:        end if
19:     end if
20:  end for
21:  if CheckAPIAccess() then
22:     input ← BufferMemory.Fetch()
23:     TransmitToLLM(input)
24:  else
25:     AssignDelay()
26:  end if
27: end while
```

We can formally represent this as (6) and (7)

$$A_{\{PDS\}} = \{Modify, Block\} \quad (6)$$

$$A_{TDP} = A_{PS} = Block \quad (7)$$

The configuration $C$ of the system can be represented as an ordered subset of $M = {PDS, TDP, PS}$: $C = (c_1, c_2, \ldots, c_k)$, where $c_i \in M$ and $k \leq n$. Each module $m_i$ has an associated function $f_i: I \to A_i \times I$, where $I$ is the input space and $A_i$ is the set of possible actions for module $m_i$. The function $f_i$ takes an input and returns both an action and a potentially modified input (in the case of PDS) or the original input (for TDP and PS). The total number of possible module combinations remains $N_{combinations} = 2^{|M|} - 1 = 7$, ranging from single-module configurations to the full set of all three modules. However, the outcome spaces for each combination are now more specifically defined. For example:

- $O(PDS) = {Modify, Block}$
- $O(TDP) = O(PS) = {Block}$
- $O(PDS, TDP) = {(Modify, Block), (Block, Block)}$

The number of possible sequences for a given combination $C$ is still determined by the number of permutations of the selected modules: $N_{sequences}(C) = |C|!$, with the total number of possible sequences across all combinations remaining at $N_{total_sequences} = 15$.

The overall system function $F$ can be defined as a composition of the individual module functions: $F(x) = f_k \circ f_{k-1} \circ \ldots \circ f_1(x)$, where $x$ is the initial input, and $\circ$ denotes function composition. This composition allows for the sequential application of safety modules in any specified order.

The decision-making process for each module can be represented as follows:

For PDS (8):

$$d_{\{PDS\}}(x) = \begin{cases} (Block, x), & \text{if } g_{\{PDS\}}(x) > t_{\{Block\}} \\ (Modify, f_{\{PDS\}}(x)) & \text{if } t_{\{Modify\}} < g_{\{PDS\}}(x) \leq t_{\{Block\}} \\ (Modify, x), & Otherwise \end{cases}$$

(8)

For TDP and PS (9):

$$d_{\{TDP|PS\}}(x) = \begin{cases} (Block, x), if \ g_{\{TDP|PS\}}(x) > t_{\{TDP|PS\}} \\ (Allow, x) \\ Otherwise \end{cases} \quad (9)$$

Where $g_i(x)$ is the classifier function for module $i$, $t_i$ are the respective thresholds, and $f_{PDS}(x)$ is the anonymization function for PDS.

The total number of possible module combinations is $N_{combinations} = 2^{|M|} - 1 = 7$, with the total number of possible sequences across all combinations being $N_{total_{sequences}} = 15$. This high number of possible configurations demonstrates the system's flexibility in adapting to various use cases and requirements.

In implementing this system, careful consideration must be given to the order of modules, particularly the placement of the PDS module. As the only module capable of modifying input, its position in the sequence can significantly impact the overall system behavior. For instance, placing PDS first ensures maximum privacy protection but may affect the performance of subsequent modules operating on anonymized data. Use cases for this system might include:

- **High-privacy scenarios**: ($PDS \rightarrow TDP \rightarrow PS$), ensuring data is anonymized before any other processing.
- **Content moderation focus**: ($TDP \rightarrow PS \rightarrow PDS$), prioritizing toxic content and prompt safety checks before privacy concerns.
- **Prompt-safety priority**: ($PS \rightarrow PDS \rightarrow TDP$), useful in scenarios where preventing prompt injections is the primary concern.

This Flexible Adaptive Sequencing mechanism, as demonstrated by the algorithm and supporting mathematical framework, offers a versatile approach to implementing layered safety measures in LLM applications. By allowing for dynamic configuration of module sequences and clearly defined module behaviors, it provides a powerful tool for addressing diverse safety, privacy, and ethical requirements in various LLM deployments.

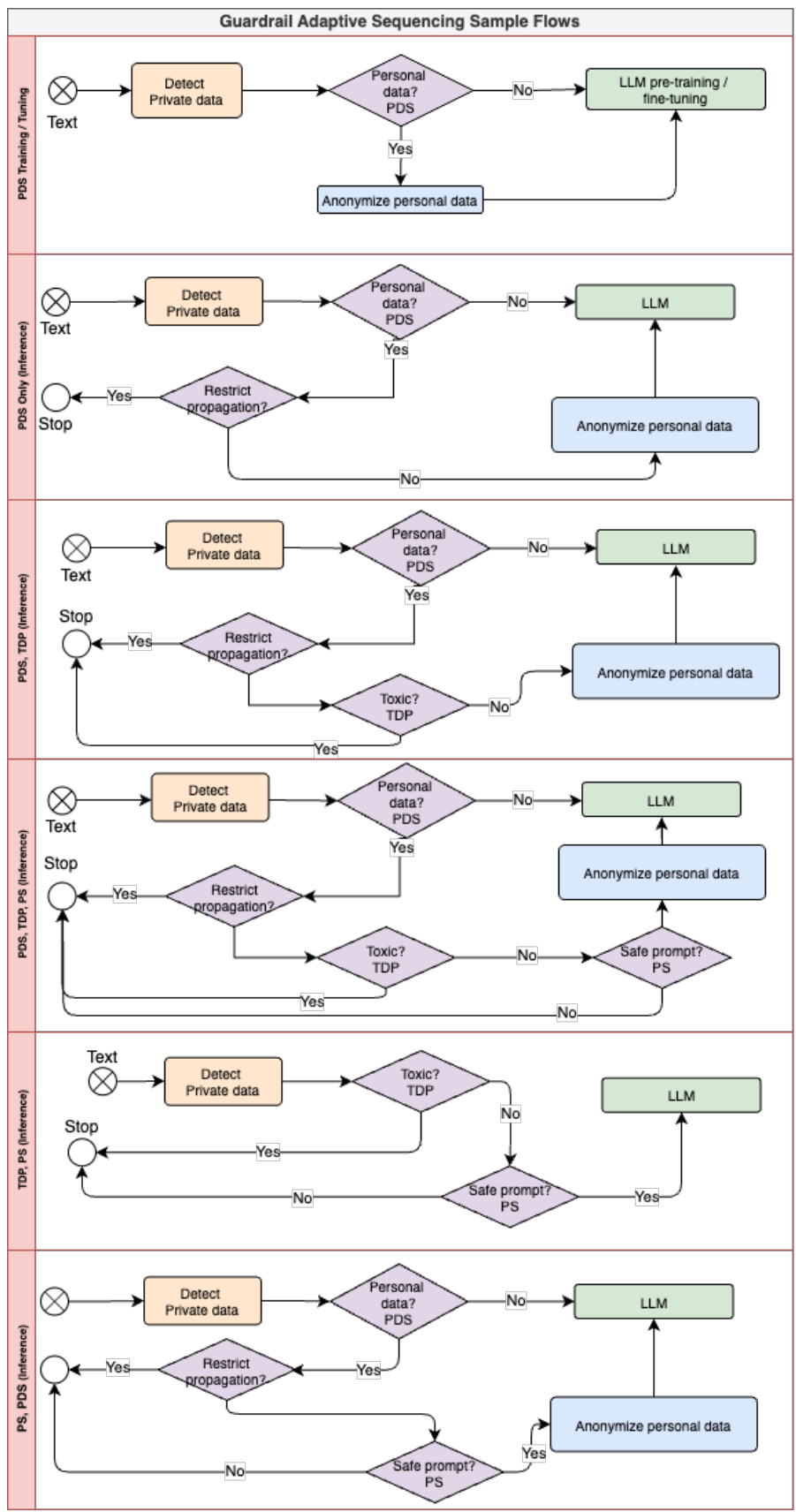

**Figure 4:** Guardrail Adaptive Sequencing of Modules - Sample Flows

## V. DISCUSSION AND FUTURE WORK

The Flexible Adaptive Sequencing mechanism, incorporating the Private Data Safety (PDS), Toxic Data Protection (TDP), and Prompt Safety (PS) modules, represents a significant step forward in ensuring trust and safety in Large Language Model (LLM) applications. This framework offers a robust and adaptable approach to addressing multiple security and ethical concerns simultaneously. However, like any complex system, it comes with both advantages and limitations that warrant discussion.

The PDS module, with its unique ability to modify or block input, provides a powerful tool for protecting user privacy. By anonymizing sensitive information before it reaches the LLM, PDS can significantly reduce the risk of personal data exposure. However, this modification of input data may potentially impact the performance of subsequent modules or the LLM itself, as the context provided by personal information is lost. Future research could explore techniques to preserve semantic meaning while anonymizing data, perhaps through the use of context-aware placeholder generation.

The TDP and PS modules, while more limited in their actions (block only), serve crucial functions in maintaining the ethical use of LLMs. TDP's ability to filter out toxic content helps create safer interaction spaces and prevents the model from generating or perpetuating harmful language. PS, on the other hand, guards against prompt injection attacks, a growing concern as LLMs become more prevalent in various applications. A potential drawback of these modules is the risk of false positives, which could lead to overly restrictive content filtering. Further work is needed to fine-tune these modules' detection algorithms to balance safety with expressiveness.

The adaptive sequencing flow is perhaps the most innovative aspect of this framework. It allows for unprecedented flexibility in applying safety measures, enabling customization for different use cases and regulatory environments. This adaptability is particularly valuable given the rapidly evolving landscape of AI ethics and regulations. However, the increased complexity of the system may pose challenges in terms of performance overhead and the need for expertise in configuring optimal module sequences. Future developments could focus on creating intelligent systems that can automatically suggest or adapt module sequences based on the specific application context and observed patterns of use.

Despite these challenges, the proposed framework provides a solid foundation for trust and safety in LLM applications. By addressing multiple aspects of security and ethics in a configurable manner, it offers a comprehensive approach that can be tailored to diverse needs. The system's ability to handle privacy, content safety, and prompt security in an integrated fashion is a significant step towards more responsible AI deployment.

Looking ahead, several avenues for future work emerge:

1. **Enhanced Module Intelligence**: Developing more sophisticated algorithms for each module, possibly incorporating advanced machine learning techniques to improve accuracy and reduce false positives.
2. **Dynamic Adaptation**: Creating systems that can dynamically adjust module sequences and parameters based on real-time performance metrics and changing contexts.
3. **Explainability and Transparency**: Implementing features that provide clear explanations for module decisions, enhancing trust and enabling more effective human oversight.
4. **Performance Optimization**: Investigating methods to reduce the computational overhead of the sequencing mechanism, particularly for high-throughput applications.
5. **Cross-lingual and Multi-modal Extensions**: Expanding the framework to handle multiple languages and various data types beyond text, such as images or audio inputs.
6. **Regulatory Alignment**: Developing tools and guidelines to help users configure the system in compliance with diverse and evolving regulatory requirements across different jurisdictions.
7. **User Interface and Experience**: Creating intuitive interfaces for configuring and monitoring the adaptive sequencing system, making it accessible to a broader range of users and organizations.

While the Flexible Adaptive Sequencing mechanism presents certain challenges and limitations, its potential for enhancing trust and safety in LLM applications is significant. By providing a flexible, comprehensive approach to handling multiple aspects of AI ethics and security, this framework lays the groundwork for more responsible and trustworthy AI systems. As research in this area progresses, we can expect to see even more sophisticated and effective methods for ensuring the safe and ethical deployment of large language models across various domains.

## VI. CONCLUSION

The rapid advancement and widespread adoption of Large Language Models (LLMs) in various domains have brought to the forefront critical concerns regarding privacy, safety, and ethical use of AI. This paper has presented a novel Flexible Adaptive Sequencing mechanism designed to address these multifaceted challenges in a comprehensive and adaptable manner. By integrating three key modules – Private Data Safety (PDS), Toxic Data Protection (TDP), and Prompt Safety (PS) – within a configurable framework, we have proposed a robust solution for enhancing trust and safety in LLM applications. The core strength of our approach lies in its flexibility and modularity. The ability to sequence and configure these safety modules in various combinations allows for tailored solutions that can meet diverse requirements across different use cases and regulatory environments. The PDS module's unique capability to modify or block input provides a powerful tool for privacy protection, while the TDP and PS modules offer critical safeguards against toxic content and malicious prompts, respectively.

Our mathematical formulation of the system, including the detailed analysis of module combinations and sequencing possibilities, provides a solid theoretical foundation for understanding and implementing this framework. The proposed algorithm, with its clear delineation of module functions and decision processes, offers a practical blueprint for real-world implementation. This combination of theoretical rigor and practical applicability positions our work as a significant contribution to the field of AI safety. The discussion of various use cases demonstrates the versatility of the Flexible Adaptive Sequencing mechanism. From high-privacy scenarios prioritizing data anonymization to content moderation focuses and prompt-safety priorities, the framework shows remarkable adaptability to different operational contexts. This flexibility is crucial in the rapidly evolving landscape of AI ethics and regulations, where one-size-fits-all solutions are increasingly inadequate.

However, we acknowledge that this framework is not without challenges. The potential impact on performance, the complexity of optimal configuration, and the need for ongoing refinement of module algorithms are areas that require further attention. The risk of false positives in content filtering and the potential loss of context through data anonymization are trade-offs that must be carefully managed. Despite these challenges, the Flexible Adaptive Sequencing mechanism represents a significant step forward in the quest for responsible AI deployment. By providing a structured yet adaptable approach to handling multiple aspects of AI ethics and security, this framework lays the groundwork for more trustworthy and safer LLM applications. The proposed future work, including enhanced module intelligence, dynamic adaptation capabilities, and improved explainability, points towards exciting avenues for further research and development.

In conclusion, as AI systems continue to grow in complexity and capability, the need for sophisticated, flexible safety mechanisms becomes increasingly critical. The Flexible Adaptive Sequencing mechanism presented in this paper offers a promising approach to meeting this need. By balancing robust protection with adaptability, it provides a foundation upon which developers, researchers, and policymakers can build to ensure that the transformative potential of LLMs is realized responsibly and ethically.

As we look to the future, the principles and methodologies outlined in this work can serve as a springboard for further innovations in AI safety. The challenge of creating AI systems that are both powerful and trustworthy is ongoing, and frameworks like the one presented here will play a crucial role in shaping the responsible development and deployment of AI technologies. Through continued research, collaboration, and ethical consideration, we can work towards a future where the benefits of AI are maximized while risks are effectively mitigated.